\documentclass{elsart}

\usepackage{graphicx,color}

\usepackage{amssymb,amsmath,bm}
\usepackage{cite}
\usepackage{url}

\newtheorem{proposition}{Proposition}

\begin{document}

\newlength\figwidth
\setlength\figwidth{0.5\columnwidth}
\newlength\imgwidth
\setlength\imgwidth{0.3\columnwidth}

\begin{frontmatter}
\title{Cryptanalysis of a computer cryptography scheme based on a filter bank}
\author[Spain]{David Arroyo\corauthref{corr}},
\author[hk-cityu]{Chengqing Li},
\author[germany]{Shujun Li} and
\author[Spain]{Gonzalo Alvarez}
\corauth[corr]{Corresponding author: David Arroyo
(david.arroyo@iec.csic.es).}
\address[Spain]{Instituto de F\'{\i}sica Aplicada, Consejo Superior de
Investigaciones Cient\'{\i}ficas, Serrano 144, 28006 Madrid,
Spain}
\address[hk-cityu]{Department of Electronic Engineering, City University of Hong Kong,
83 Tat Chee Avenue, Kowloon Tong, Hong Kong SAR, China}
\address[germany]{FernUniversit\"{a}t in Hagen, Lehrgebiet Informationstechnik, Universit\"{a}tsstra{\ss}e 27, 58084 Hagen, Germany}

\begin{abstract}
This paper analyzes the security of a recently-proposed signal
encryption scheme based on a filter bank. A very critical weakness
of this new signal encryption procedure is exploited in order to
successfully recover the associated secret key.
\end{abstract}
\end{frontmatter}

\section{Introduction}

The application of chaotic systems to cryptographical issues has
been a very important research topic since the 1990s
\cite{liThesis,Alvarez:Survey:ICCST99,kocarev2001a,yang04}. This
interest was motivated by the close similarities between some
properties of chaotic systems and some characteristics of
well-designed cryptosystems \cite[Table~1]{Alvarez06a}.
Nevertheless, there exist security defects in some chaos-based
cryptosystems such that they can be partially or totally broken
\cite{Alvarez:BreakingHenon:CSF2004,Alvarez:BreakingCPM:CSF2004,Alvarez:Improved:CSF2004,AlvarezLi:BreakingPS:2005,Li:BreakingBuWang:CSF2005,Alvarez:DeniableAuthentication:CSF2005}.

In \cite{ling07} the encryption procedure is carried out by
decomposing the input plaintext signal into two different subbands
and masking each of them with a pseudo-random number sequence
generated by iterating the chaotic logistic map. The decomposition
of the input plaintext signal $x[n]$ is driven by

\begin{eqnarray}
t_0[n]&=&K_0\sum_{\forall m}x[m]h_0[2n-m],\\
t_1[n]&=&K_1\sum_{\forall m}x[m]h_1[2n-m],
\end{eqnarray}
where $h_0, h_1$ are so-called ``analysis filters'' and $K_0$, $K_1$
are gain factors.

Then, the masking stage generates the ciphertext signal
$(v_0[n],v_1[n])$ according to the following equations:
\begin{eqnarray}
v_0[n] &=& t_0[n]+\alpha_0(t_1[n]),\label{equation:v0}\\
v_1[n] &=& t_1[n] -\alpha_1(v_0[n]),\label{equation:v1}
\end{eqnarray}
where $\alpha_i(u)=u+s_i[n]$ and $s_i[n]$ is the state variable of a
logistic map with control parameter $\lambda_i\in(3,4)$ defined as
follows\footnote{In \cite{ling07}, the authors use $x_i$ to denote
the state variable of the logistic map. However, this nomenclature
may cause confusion because the plaintext signal is denoted by $x$.
Therefore, we turn to use another letter, $s$. In addition, we unify
the representation of $x_i(k)$ to be in the form $s_i[n]$ because
all other signals are in the latter form.}
\begin{equation}
s_i[n]=\lambda_i s_i[n-1](1-s_i[n-1]).\label{equation:alfa}
\end{equation}
Substituting $\alpha_i(u)=u+s_i[n]$ into Eqs.~(\ref{equation:v0})
and (\ref{equation:v1}), we have
\begin{eqnarray}
v_0[n] &=& (t_0[n]+t_1[n])+s_0[n],\label{equation:v0b}\\
v_1[n] &=& (t_1[n]-v_0[n])-s_1[n].\label{equation:v1b}
\end{eqnarray}

The secret key of the cryptosystem is composed of the initial
conditions and the control parameters of the two logistic maps
involved, i.e., $s_0[0]$, $s_1[0]$, $\lambda_0$ and $\lambda_1$.

The decryption procedure is carried out by doing
\begin{eqnarray}
    t_1[n] &=& v_1[n] + \alpha_1(v_0[n]),\\
    t_0[n] &=& v_0[n] - \alpha_0(t_1[n]).
\end{eqnarray}
Then, the plaintext signal is recovered with the following filtering
operations:
\begin{equation}
\tilde{x}[n]=\frac{1}{K_0}\sum_{\forall
m}t_0[m]f_0[n-2m]+\frac{1}{K_1}\sum_{\forall m}t_1[m]f_1[n-2m],
\end{equation}
where $f_0,f_1$ are so-called ``synthesis filters''. To ensure the
correct recovery of the plaintext signal, the analysis and synthesis
filters must satisfy a certain requirement as shown in Eq.~(8) of
\cite{ling07}. The reader is referred to \cite{ling07} for more
information about the inner working of the cryptosystem.

This paper focuses on the security analysis of the above
cryptosystem. The next section points out a security problem about
the reduction of the key space. Section~\ref{section:attack}
discusses how to recover the secret key of the cryptosystem by a
known-plaintext attack. In the last section the conclusion is given.

\section{Reduction of the key space}
\label{section:considerations}

\begin{figure}
    \centering
    \includegraphics{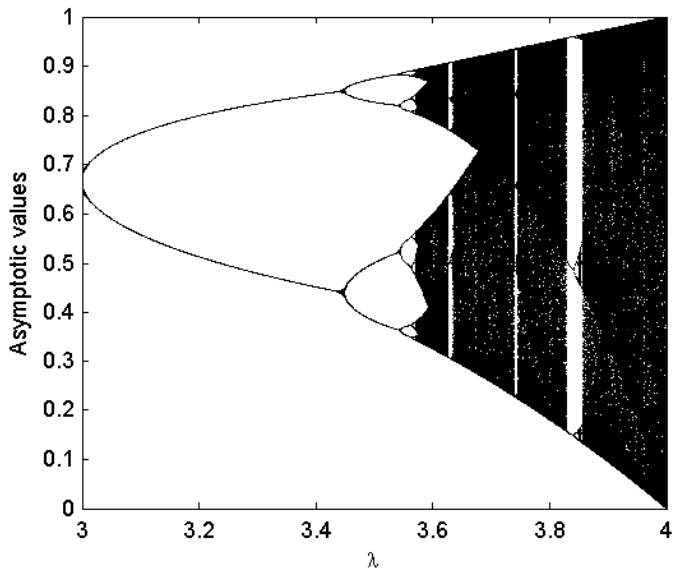}
    \caption{Bifurcation diagram of the logistic map}
    \label{figure:logBif}
\end{figure}

As it is pointed out in \cite[Rule 5]{Alvarez06a}, the key related
to a chaotic cryptosystem should avoid non-chaotic areas. In
\cite{ling07} it is claimed that the key space of the cryptosystem
under study is given by the set of values $\lambda_i$ and $s_i[0]$
satisfying $3<\lambda_i<4$ and $0<s_i[0]<1$ for $i=0,1$. However,
when looking at the bifurcation diagram of the logistic map (Fig.
\ref{figure:logBif}), it is obvious that not all candidate values of
$\lambda_i$ and $s_i[0]$ are valid to ensure the chaoticity of the
logistic map. There are periodic windows which have to be avoided by
carefully choosing $\lambda_i$. As a consequence, the available key
space is drastically reduced.

\section{Known-plaintext attack}
\label{section:attack}

In a known-plaintext attack the cryptanalyst possesses a plaintext
signal $\{x[n]\}$ and its corresponding encrypted subband signals
$\{v_0[n]\}$ and $\{v_1[n]\}$. Because $\{h_0[n]\}$, $\{h_1[n]\}$,
$K_0$ and $K_1$ are public, we can get $\{t_0[n]\}$ and $\{t_1[n]\}$
from $\{x[n]\}$. Then we can get the values of $\{s_0[n]\}$ and
$\{s_1[n]\}$ as follows:
\begin{eqnarray}
    s_0[n] & = & v_0[n]-t_0[n]-t_1[n],\\
    s_1[n] & = & t_1[n]-v_0[n]-v_1[n].
\end{eqnarray}

For $n=0$, the values of the subkeys $s_0[0]$ and $s_1[0]$ have been
obtained. Furthermore, we can obtain the control parameters by just
doing the following operations for $i=0,1$:
\begin{equation}
\lambda_i=\frac{s_i[n+1]}{s_i[n](1-s_i[n])}.\label{eq:RecoverParameter}
\end{equation}

In \cite{ling07}, the authors did not give any discussion about the
finite precision about the implementation of the cryptosystem in
computers. If the floating-point precision is used, then the value
of $\lambda_i$ can be estimated very accurately. It was
experimentally verified that the error for the estimation of
$\lambda_i$ using (\ref{eq:RecoverParameter}), and working with
floating-point precision, was never greater that $4\cdot 10^{-12}$.
If the fixed-point precision is adopted, the deviation of the
parameter $\lambda_i$ estimated exploiting
Eq.~(\ref{eq:RecoverParameter}) from the real $\lambda_i$ may be
very large. Fortunately, according to the following Proposition
\ref{proposition} \cite[Proposition 2]{Li:AttackingRCES2007}, the
error is limited to $2^4/2^L$ (which means only $2^4$ possible
candidate values to be further guessed) when $s[n+1]\geq 0.5$.

\begin{proposition}\label{proposition}
Assume that the logistic map $s[n+1]=\lambda\cdot s[n]\cdot(1-s[n])$
is iterated with $L$-bit fixed-point arithmetic and that $s(n+1)\geq
2^{-i}$, where $1\leq i\leq L$. Then, the following inequality
holds: $|\lambda-\widetilde{\lambda}|\leq 2^{i+3}/2^L$, where
$\widetilde{\lambda}=\dfrac{s[n+1]}{s[n]\cdot(1-s[n])}$.
\end{proposition}

\section{Conclusion}
\label{section:conclusions}

In this paper we have analyzed the security properties of the
cryptosystem proposed in \cite{ling07}. It has been shown that there
exists a great number of weak keys derived from the fact that the
logistic map is not always chaotic. In addition, the cryptosystem is
very weak against a known-plaintext attack in the sense that the
secret key can be totally recovered using a very short plaintext.
Consequently, the cryptosystem introduced by \cite{ling07} should be
discarded as a secure way of exchanging information.

\section*{Acknowledgments}
The work described in this paper was partially supported by
Minis\-terio de Educaci\'on y Ciencia of Spain, Research Grant
SEG2004-02418. Shujun Li was supported by the Alexander von Humboldt
Foundation, Germany.

\bibliographystyle{elsart-num}
\bibliography{database}

\begin{thebibliography}{10}
\expandafter\ifx\csname url\endcsname\relax
  \def\url#1{\texttt{#1}}\fi
\expandafter\ifx\csname urlprefix\endcsname\relax\def\urlprefix{URL }\fi

\bibitem{liThesis}
S.~Li, Analyses and new designs of digital chaotic ciphers, Ph.D. thesis,
  School of Electronic and Information Engineering, Xi'an Jiaotong University,
  Xi'an, China, available online at \url{http://www.hooklee.com/pub.html} (June
  2003).

\bibitem{Alvarez:Survey:ICCST99}
G.~Alvarez, F.~Montoya, M.~Romera, G.~Pastor, Chaotic cryptosystems, in: L.~D.
  Sanson (Ed.), Proc. 33rd Annual 1999 International Carnahan Conference on
  Security Technology, IEEE, 1999, pp. 332--338.

\bibitem{kocarev2001a}
L.~Kocarev, Chaos-based cryptography: A brief overview, IEEE Circuits Syst.
  Mag. 1 (2001) 6--21.

\bibitem{yang04}
T.~Yang, A survey of chaotic secure communication systems, Int. J. Comp.
  Cognition 2~(2) (2004) 81--130.

\bibitem{Alvarez06a}
G.~Alvarez, S.~Li, Some basic cryptographic requirements for chaos-based
  cryptosystems, International Journal of Bifurcation and Chaos 16~(8) (2006)
  2129--2151.

\bibitem{Alvarez:BreakingHenon:CSF2004}
G.~Alvarez, F.~Montoya, M.~Romera, G.~Pastor, Cryptanalyzing a discrete-time
  chaos synchronization secure communication system, Chaos, Solitons and
  Fractals 21~(3) (2004) 689--694.

\bibitem{Alvarez:BreakingCPM:CSF2004}
G.~Alvarez, F.~Montoya, M.~Romera, G.~Pastor, Breaking parameter modulated
  chaotic secure communication system, Chaos, Solitons and Fractals 21~(4)
  (2004) 783--787.

\bibitem{Alvarez:Improved:CSF2004}
G.~Alvarez, F.~Montoya, M.~Romera, G.~Pastor, Cryptanalyzing an improved
  security modulated chaotic encryption scheme using ciphertext absolute value,
  Chaos, Solitons and Fractals 23~(5) (2004) 1749--1756.

\bibitem{AlvarezLi:BreakingPS:2005}
G.~Alvarez, S.~Li, F.~Montoya, G.~Pastor, M.~Romera, Breaking projective chaos
  synchronization secure communication using filtering and generalized
  synchronization, Chaos, Solitons and Fractals 24~(3) (2005) 775--783.

\bibitem{Li:BreakingBuWang:CSF2005}
S.~Li, G.~Alvarez, G.~Chen, Breaking a chaos-based secure communication scheme
  designed by an improved modulation method, Chaos, Solitons and Fractals
  25~(1) (2005) 109--120.

\bibitem{Alvarez:DeniableAuthentication:CSF2005}
G.~Alvarez, Security problems with a chaos-based deniable authentication
  scheme, Chaos, Solitons and Fractals 26~(1) (2005) 7--11.

\bibitem{ling07}
B.~W.-K. Ling, C.~Y.-F. Ho, P.~K.-S. Tam, Chaotic filter bank for computer
  cryptography, Chaos, Solitons and Fractals 34 (2007) 817--824.

\bibitem{Li:AttackingRCES2007}
S.~Li, C.~Li, G.~Chen, K.-T. Lo, Cryptanalysis of the {RCES/RSES} image
  encryption scheme, J. Systems and Software, in press,
  doi:10.1016/j.jss.2007.07.037 (2007).

\end{thebibliography}
\end{document}